\documentclass{stacs_proc}

\usepackage{latexsym,amsbsy,amssymb,amsmath,amsfonts}
\usepackage{euscript}

\usepackage{graphicx}
\usepackage{color}

\newcommand{\opstyle}[1]{\mathrm{#1}}
\newcommand{\cNL}{\opstyle{NL}}
\newcommand{\ccoNL}{\opstyle{coNL}}
\newcommand{\cPSPACE}{\opstyle{PSPACE}}
\newcommand{\cDSPACE}{\opstyle{DSPACE}}

\newcommand\co{\mathord{\mbox{\rm co}}}
\newcommand{\bc}[1]{{\mathrm{BC}(#1)}}

\newcommand\lsig[1]{\Sigma^{\varrho}_{#1}}   
\newcommand\sigd{\Sigma^{\rm\tau_d}}
\newcommand\sigt{\Sigma^{\rm\tau}}
\newcommand\sig[1]{\Sigma^\sigma_{#1}}
\newcommand\bsig[1]{\sig{#1}} 

\newcommand\bbc[1]{\bc{\bsig{#1}}}
\newcommand\lbc[1]{\bc{\lsig{#1}}}

\newcommand\lpig[1]{\Pi^{\varrho}_{#1}}
\newcommand\pigd{\Pi^{\rm\tau_d}}
\newcommand\pigt{\Pi^{\rm\tau}}
\newcommand\pig[1]{\Pi^\sigma_{#1}}
\newcommand\bpig[1]{\pig{#1}}

\newcommand{\cond}{\,\big|\,}

\newcommand{\mymod}{\mathrm{mod\ }}
\newcommand{\az}{{\Sigma}}
\newcommand{\sowp}{{\az^+}}
\newcommand{\alf}[1]{{\alpha(#1)}}      

\newcommand{\sw}{{\preceq}} 
\newcommand{\gdw}{\,\mathop{\Leftrightarrow}\,}
\newcommand{\isdefinedl}{\mathop{=}\limits^{\mbox{%
    \raisebox{-0.15ex}[0ex][0ex]{$\scriptscriptstyle df$}}}}
\newcommand{\isdefined}{\isdefinedl}
\newcommand\eqdef{\isdefined}

\newcommand{\con}[2][M]{\mathop{\longrightarrow}\limits_{#1}^{#2}}

\newcommand\eps{\varepsilon}             
\newcommand{\mf}[1]{{\mathcal{#1}}}         
\newcommand{\unmark}[1]{\overline{#1}}  
\newcommand\set[2]{\bigl\{\,#1\bigm| #2\,\bigr\}}

\newcommand{\FO}{\opstyle{FO}}

\begin{document}

\title[Efficient Algorithms for Boolean Hierarchies of Regular Languages]
					{Efficient Algorithms for Membership in\\ 
					 Boolean Hierarchies of Regular Languages}

\author[lab1]{C. Gla{\ss}er}{C. Gla{\ss}er}
\address[lab1]{Universit{\"a}t W{\"u}rzburg, Germany.}
\email{glasser@informatik.uni-wuerzburg.de}  

\author[lab2]{H. Schmitz}{H. Schmitz}
\address[lab2]{Fachhochschule Trier, Germany.}    
\email{schmitz@informatik.fh-trier.de}  

\author[lab3]{V. Selivanov}{V. Selivanov}
\address[lab3]{A.~P.~Ershov Institute of Informatics Systems, Russia.}    
\email{vseliv@nspu.ru}  

\thanks{This work was done during a stay of the third author at the
    University of W\"urzburg, supported  by  DFG Mercator program  and by RFBR grant 07-01-00543a.} 

\keywords{automata and formal languages, computational complexity,
		dot-depth hier\-archy, Boolean hierarchy, decidability, efficient algorithms}


\begin{abstract}
  \noindent The purpose of this paper is to provide {\em efficient} algorithms that decide membership for classes
    of several Boolean hierarchies for which efficiency (or even decidability) were previously not known.
    We develop new forbidden-chain characterizations for the single levels of these hierarchies and
    obtain the following results:
    \begin{itemize}
        \item The classes of the Boolean hierarchy over level $\Sigma_1$
                of the dot-depth hierarchy are decidable in $\cNL$
                (previously only the decidability was known).
                The same remains true if predicates mod $d$ for fixed $d$ are allowed.
        \item If modular predicates for arbitrary $d$ are allowed, then
                the classes of the Boolean hierarchy over level $\Sigma_1$ are decidable.
        \item For the restricted case of a two-letter alphabet,
            the classes of the Boolean hierarchy over level $\Sigma_2$
        of the Straubing-Th\'{e}rien hierarchy are decidable in $\cNL$.
        This is the first decidability result for this hierarchy.
        \item The membership problems for all mentioned Boolean-hierarchy classes are logspace many-one hard for $\cNL$.
        \item The membership problems for quasi-aperiodic languages and for
                $d$-quasi-aperiodic languages are logspace many-one complete for $\cPSPACE$.
    \end{itemize}
\end{abstract}

\maketitle

\stacsheading{2008}{337-348}{Bordeaux}
\firstpageno{337}

\section*{Introduction}\label{S:one}

The study of decidability and complexity questions for classes of regular languages
is a central research topic in automata theory.
Its importance stems from the fact that finite automata are fundamental
to many branches of computer science, e.g.,
databases, operating systems, verification, hardware and software design.

There are many examples for decidable classes of regular languages (e.g., locally testable languages),
while the decidability of other classes is still a challenging open question (e.g., dot-depth two, generalized star-height).
Moreover, among the decidable classes there is a broad range of complexity results. For some of them, e.g.,
the class of piecewise testable languages, efficient algorithms are known that work in nondeterministic logarithmic space ($\cNL$)
and hence in polynomial time. For other classes, a membership test needs more resources, e.g.,
deciding the membership in the class of star-free languages is $\cPSPACE$-complete.

The purpose of this paper is to provide {\em efficient} algorithms that decide membership for classes
of several Boolean hierarchies for which efficiency (or even decidability) were not previously known.
Many of the known efficient decidability results for classes of regular languages are based on so-called
forbidden-pattern characterizations. Here a language belongs to a class of regular languages if and only if
its deterministic finite automaton does {\em not} have a certain subgraph (the forbidden pattern)
in its transition graph. Usually, such a condition can be checked efficiently, e.g., in nondeterministic logarithmic space
\cite{stern85,cpp93,gs00,gs00b}.

However, for the Boolean hierarchies considered in this paper, the design of efficient algorithm is more involved,
since here no forbidden-pattern characterizations are known.
More precisely, wherever decidability is known, it is obtained from a characterization of the corresponding class
in terms of {\em forbidden} alternating chains of word extensions.
Though the latter also is a forbidden property, the known characterizations are not efficiently
checkable in general. (Exceptions are the special `local' cases $\lsig{1}(n)$ and ${\mathcal C}^1_k(n)$
where decidability in $\cNL$ is known \cite{scwa98a,sch01}.)
To overcome these difficulties, we first develop alternative forbidden-chain characterizations
(they essentially ask only for certain reachability conditions in transition graphs).
From our new characterizations we obtain efficient algorithms for membership tests in $\cNL$.
For two of the considered Boolean hierarchies, these are the first decidable characterizations at all,
i.e., for the classes $\lsig{2}(n)$ for the alphabet $A=\{a,b\}$, and for the classes $\sigt_1(n)$).

\begin{figure}[t]
    \centering
\begin{picture}(0,0)%
\includegraphics{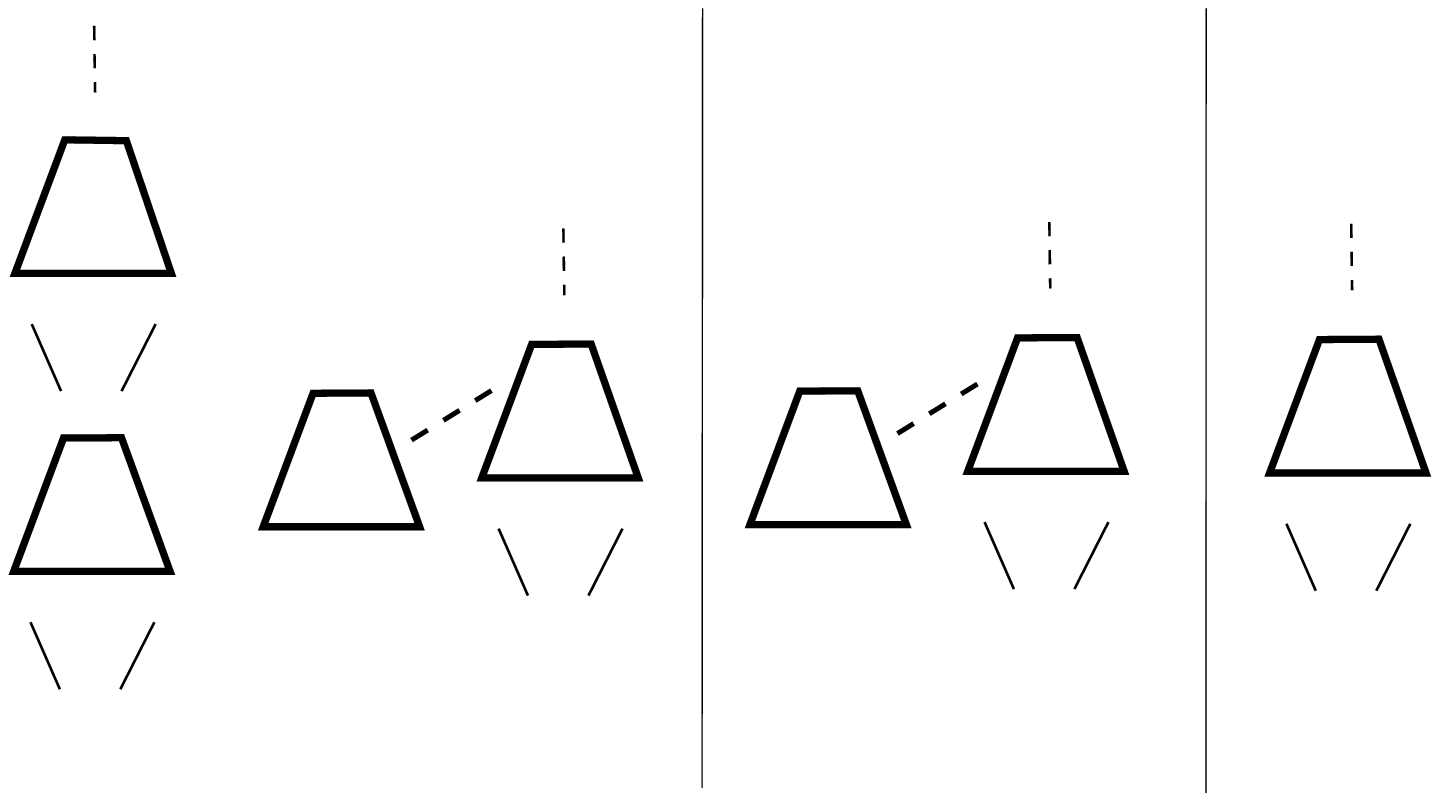}%
\end{picture}%
\setlength{\unitlength}{2960sp}%
\begingroup\makeatletter\ifx\SetFigFont\undefined%
\gdef\SetFigFont#1#2#3#4#5{%
  \reset@font\fontsize{#1}{#2pt}%
  \fontfamily{#3}\fontseries{#4}\fontshape{#5}%
  \selectfont}%
\fi\endgroup%
\begin{picture}(9269,5118)(463,-6664)
\put(749,-3162){\makebox(0,0)[lb]{\smash{{\SetFigFont{9}{10.8}{\rmdefault}{\mddefault}{\updefault}{\color[rgb]{0,0,0}$\{\lsig{2}(n)\}$}%
}}}}
\put(742,-5038){\makebox(0,0)[lb]{\smash{{\SetFigFont{9}{10.8}{\rmdefault}{\mddefault}{\updefault}{\color[rgb]{0,0,0}$\{\lsig{1}(n)\}$}%
}}}}
\put(478,-5371){\makebox(0,0)[lb]{\smash{{\SetFigFont{9}{10.8}{\rmdefault}{\mddefault}{\updefault}{\color[rgb]{0,0,0}$\lsig{1}$}%
}}}}
\put(796,-4238){\makebox(0,0)[lb]{\smash{{\SetFigFont{9}{10.8}{\rmdefault}{\mddefault}{\updefault}{\color[rgb]{0,0,0}$\bc{\lsig{1}}$}%
}}}}
\put(1510,-5371){\makebox(0,0)[lb]{\smash{{\SetFigFont{9}{10.8}{\rmdefault}{\mddefault}{\updefault}{\color[rgb]{0,0,0}$\lpig{1}$}%
}}}}
\put(734,-6155){\makebox(0,0)[lb]{\smash{{\SetFigFont{9}{10.8}{\rmdefault}{\mddefault}{\updefault}{\color[rgb]{0,0,0}$\lsig{0}=\lpig{0}$}%
}}}}
\put(486,-3463){\makebox(0,0)[lb]{\smash{{\SetFigFont{9}{10.8}{\rmdefault}{\mddefault}{\updefault}{\color[rgb]{0,0,0}$\lsig{2}$}%
}}}}
\put(804,-2330){\makebox(0,0)[lb]{\smash{{\SetFigFont{9}{10.8}{\rmdefault}{\mddefault}{\updefault}{\color[rgb]{0,0,0}$\bc{\lsig{2}}$}%
}}}}
\put(1518,-3463){\makebox(0,0)[lb]{\smash{{\SetFigFont{9}{10.8}{\rmdefault}{\mddefault}{\updefault}{\color[rgb]{0,0,0}$\lpig{2}$}%
}}}}
\put(3738,-4439){\makebox(0,0)[lb]{\smash{{\SetFigFont{9}{10.8}{\rmdefault}{\mddefault}{\updefault}{\color[rgb]{0,0,0}$\{\bsig{1}(n)\}$}%
}}}}
\put(3474,-4772){\makebox(0,0)[lb]{\smash{{\SetFigFont{9}{10.8}{\rmdefault}{\mddefault}{\updefault}{\color[rgb]{0,0,0}$\bsig{1}$}%
}}}}
\put(3792,-3639){\makebox(0,0)[lb]{\smash{{\SetFigFont{9}{10.8}{\rmdefault}{\mddefault}{\updefault}{\color[rgb]{0,0,0}$\bc{\bsig{1}}$}%
}}}}
\put(4506,-4772){\makebox(0,0)[lb]{\smash{{\SetFigFont{9}{10.8}{\rmdefault}{\mddefault}{\updefault}{\color[rgb]{0,0,0}$\bpig{1}$}%
}}}}
\put(3730,-5556){\makebox(0,0)[lb]{\smash{{\SetFigFont{9}{10.8}{\rmdefault}{\mddefault}{\updefault}{\color[rgb]{0,0,0}$\bsig{0}=\bpig{0}$}%
}}}}
\put(5477,-4749){\makebox(0,0)[lb]{\smash{{\SetFigFont{9}{10.8}{\rmdefault}{\mddefault}{\updefault}{\color[rgb]{0,0,0}$\{\mathcal{C}^d_k(n)\}$}%
}}}}
\put(6840,-5514){\makebox(0,0)[lb]{\smash{{\SetFigFont{9}{10.8}{\rmdefault}{\mddefault}{\updefault}{\color[rgb]{0,0,0}$\sigd_0=\pigd_0$}%
}}}}
\put(7616,-4730){\makebox(0,0)[lb]{\smash{{\SetFigFont{9}{10.8}{\rmdefault}{\mddefault}{\updefault}{\color[rgb]{0,0,0}$\pigd_1$}%
}}}}
\put(6902,-3597){\makebox(0,0)[lb]{\smash{{\SetFigFont{9}{10.8}{\rmdefault}{\mddefault}{\updefault}{\color[rgb]{0,0,0}$\bc{\sigd_1}$}%
}}}}
\put(6584,-4730){\makebox(0,0)[lb]{\smash{{\SetFigFont{9}{10.8}{\rmdefault}{\mddefault}{\updefault}{\color[rgb]{0,0,0}$\sigd_1$}%
}}}}
\put(5508,-3938){\makebox(0,0)[lb]{\smash{{\SetFigFont{9}{10.8}{\rmdefault}{\mddefault}{\updefault}{\color[rgb]{0,0,0}$\bc{\mathcal{C}^d_k}$}%
}}}}
\put(5187,-5111){\makebox(0,0)[lb]{\smash{{\SetFigFont{9}{10.8}{\rmdefault}{\mddefault}{\updefault}{\color[rgb]{0,0,0}$\mathcal{C}^d_k$}%
}}}}
\put(6158,-5132){\makebox(0,0)[lb]{\smash{{\SetFigFont{9}{10.8}{\rmdefault}{\mddefault}{\updefault}{\color[rgb]{0,0,0}$\co\mathcal{C}^d_k$}%
}}}}
\put(6809,-4430){\makebox(0,0)[lb]{\smash{{\SetFigFont{9}{10.8}{\rmdefault}{\mddefault}{\updefault}{\color[rgb]{0,0,0}$\{\sigd_1(n)\}$}%
}}}}
\put(3043,-5146){\makebox(0,0)[lb]{\smash{{\SetFigFont{9}{10.8}{\rmdefault}{\mddefault}{\updefault}{\color[rgb]{0,0,0}$\co\mathcal{C}^1_k$}%
}}}}
\put(8780,-4408){\makebox(0,0)[lb]{\smash{{\SetFigFont{9}{10.8}{\rmdefault}{\mddefault}{\updefault}{\color[rgb]{0,0,0}$\{\sigt_1(n)\}$}%
}}}}
\put(8516,-4741){\makebox(0,0)[lb]{\smash{{\SetFigFont{9}{10.8}{\rmdefault}{\mddefault}{\updefault}{\color[rgb]{0,0,0}$\sigt_1$}%
}}}}
\put(8834,-3608){\makebox(0,0)[lb]{\smash{{\SetFigFont{9}{10.8}{\rmdefault}{\mddefault}{\updefault}{\color[rgb]{0,0,0}$\bc{\sigt_1}$}%
}}}}
\put(9548,-4741){\makebox(0,0)[lb]{\smash{{\SetFigFont{9}{10.8}{\rmdefault}{\mddefault}{\updefault}{\color[rgb]{0,0,0}$\pigt_1$}%
}}}}
\put(8772,-5525){\makebox(0,0)[lb]{\smash{{\SetFigFont{9}{10.8}{\rmdefault}{\mddefault}{\updefault}{\color[rgb]{0,0,0}$\sigt_0=\pigt_0$}%
}}}}
\put(2072,-5125){\makebox(0,0)[lb]{\smash{{\SetFigFont{9}{10.8}{\rmdefault}{\mddefault}{\updefault}{\color[rgb]{0,0,0}$\mathcal{C}^1_k$}%
}}}}
\put(2362,-4763){\makebox(0,0)[lb]{\smash{{\SetFigFont{9}{10.8}{\rmdefault}{\mddefault}{\updefault}{\color[rgb]{0,0,0}$\{\mathcal{C}^1_k(n)\}$}%
}}}}
\put(2393,-3952){\makebox(0,0)[lb]{\smash{{\SetFigFont{9}{10.8}{\rmdefault}{\mddefault}{\updefault}{\color[rgb]{0,0,0}$\bc{\mathcal{C}^1_k}$}%
}}}}
\put(2594,-1744){\makebox(0,0)[lb]{\smash{{\SetFigFont{9}{10.8}{\rmdefault}{\mddefault}{\updefault}{\color[rgb]{0,0,0}aperiodic}%
}}}}
\put(3782,-6579){\makebox(0,0)[lb]{\smash{{\SetFigFont{9}{10.8}{\rmdefault}{\mddefault}{\updefault}{\color[rgb]{0,0,0}DDH ($A^+$)}%
}}}}
\put(827,-6589){\makebox(0,0)[lb]{\smash{{\SetFigFont{9}{10.8}{\rmdefault}{\mddefault}{\updefault}{\color[rgb]{0,0,0}STH ($A^*$)}%
}}}}
\put(1024,-2622){\makebox(0,0)[lb]{\smash{{\SetFigFont{9}{10.8}{\rmdefault}{\mddefault}{\updefault}{\color[rgb]{0,0,0}for}%
}}}}
\put(5869,-1764){\makebox(0,0)[lb]{\smash{{\SetFigFont{9}{10.8}{\rmdefault}{\mddefault}{\updefault}{\color[rgb]{0,0,0}$d$-quasi-aperiodic}%
}}}}
\put(812,-2887){\makebox(0,0)[lb]{\smash{{\SetFigFont{9}{10.8}{\rmdefault}{\mddefault}{\updefault}{\color[rgb]{0,0,0}$|A|\!=\!2$}%
}}}}
\put(8627,-1775){\makebox(0,0)[lb]{\smash{{\SetFigFont{9}{10.8}{\rmdefault}{\mddefault}{\updefault}{\color[rgb]{0,0,0}quasi-aperiodic}%
}}}}
\end{picture}%
    	\caption{Boolean hierarchies considered in this paper.}
    \label{pap_b1}
\end{figure}

{\bf Definitions. }
We sketch the definitions of the Boolean hierarchies considered in this paper.
$\lsig{1}$ denotes the class of languages definable by
first-order $\Sigma_1$-sentences over the signature $\varrho=\{\leq,Q_a,\ldots\}$
where for every letter $a \in A$, $Q_a(i)$ is true if and only if
the letter $a$ appears at the $i$-th position in the word.
$\lsig{1}$ equals level $1/2$ of the Straubing-Th\'{e}rien hierarchy (STH for short) \cite{str81,the81,str85,pepi86}.
$\lsig{2}$ is the class of languages definable by similar first-order $\Sigma_2$-sentences;
this class equals level $3/2$ of the Straubing-Th\'{e}rien hierarchy. 
Let $\sigma$ be the signature obtained from $\varrho$ by adding constants for the
minimum and maximum positions in words and adding functions that compute
the successor and the predecessor of positions.
$\bsig{1}$ denotes the class of languages definable by
first-order $\Sigma_1$-sentences of the signature $\sigma$;
this class equals level $1/2$ of the dot-depth hierarchy (DDH for short) \cite{cobro71,tho82}.
Let $\tau_d$ be the signature obtained from $\sigma$ by adding the unary predicates
$P_d^0,\ldots,P_d^{d-1}$ where $P_d^j(i)$ is true if and only if $i \equiv j (\mymod d)$.
Let $\tau$ be the union of all $\tau_d$.
$\sigd_1$ (resp., $\sigt_1$) is the class of languages definable by
first-order $\Sigma_1$-sentences of the signature $\tau_d$ (resp., $\tau$).
${\mathcal C}^d_k$ is the generalization of $\lsig{1}$ where neighborhoods of
$k+1$ consecutive letters and distances modulo $d$ are expressible (Definition~\ref{pap_kd}).
For a class $\mathcal{D}$ (in our case one of the classes $\lsig{1}$, $\bsig{1}$,
${\mathcal C}^d_k$, $\sigd_1$, $\sigt_1$, and $\lsig{2}$ for $|A|=2$),
the {\em Boolean hierarchy over $\mathcal{D}$} is the family of classes
$$\mathcal{D}(n) \isdefined \{L \cond L = L_1 - (L_2-(\ldots-L_n)) \mbox{ where } 
L_1, \ldots, L_n \in \mathcal{D} \mbox{ and } L_1 \supseteq L_2 \supseteq \cdots \supseteq L_n\}.$$
The Boolean hierarchies considered in this paper are illustrated in Figure~\ref{pap_b1}.

{\bf Our Contribution. }
The paper contributes to the understanding of Boolean hierarchies of regular languages in two ways:
\begin{enumerate}
    \item For the classes $\bsig{1}(n)$, $\sigd_1(n)$, and $\lsig{2}(n)$ for the alphabet $A=\{a,b\}$
    we prove new characterizations in terms of forbidden alternating chains.
    In case of $\lsig{2}(n)$ for the alphabet $A=\{a,b\}$, this is the first characterization
    of this class.
    \item For the classes $\bsig{1}(n)$, ${\mathcal C}^d_k(n)$, $\sigd_1(n)$, and $\lsig{2}(n)$ for the alphabet $A=\{a,b\}$
    we construct the first efficient algorithms for testing membership in these classes.
    In particular, this yields the decidability of the classes
    $\sigt_1(n)$,  and of $\lsig{2}(n)$ for the alphabet $A=\{a,b\}$.
\end{enumerate}

We also show that the membership problems for all mentioned Boolean-hierarchy classes are logspace many-one hard for $\cNL$.
An overview of the obtained decidability and complexity results can be found in Table~1. 
Moreover, we prove that the membership problems
for quasi-aperiodic languages and for
$d$-quasi-aperiodic languages are logspace many-one complete for $\cPSPACE$.

Boolean hierarchies can also be seen as fine-grain measures for
regular languages in terms of descriptional complexity. Note that the
Boolean hierarchies considered in this paper do not collapse \cite{sh98,ss00,s04}.
Moreover, all these hierarchies either are known or turn out to be
decidable (see Table~1 
for the attribution of these results).
If in addition the Boolean closure of the base class is decidable, then we can even exactly
compute the Boolean level of a given language.
By known results (summarized in Theorem~\ref{pap_thm_decidibility_bc}),
one can do this exact computation of the level for the Boolean hierarchies over
$\lsig{1}$, $\lsig{2}$ (for alphabet $A=\{a,b\}$), ${\mathcal C}^1_k$, $\bsig{1}$, and $\sigt_1$.
To achieve the same for the Boolean hierarchies over ${\mathcal C}^d_k$ and $\sigd_1$
we need the decidability of their Boolean closures which is not known.

{\bf Related Work. } Due to the many characterizations of regular
languages there are several approaches to attack decision problems
on subclasses of regular languages: Among them there is the
algebraic, the automata-theoretic, and the logical approach. In this
paper we mainly use the logical approach which has a long tradition
starting with the early work of Trakhtenbrot \cite{tra58} and
B\"uchi \cite{bue60}.
 Decidability questions for Boolean
hierarchies over classes of concatenation hierarchies were
previously studied by \cite{scwa98a,sch01,gs01b,s04}. Enrichments of
the first-order logics related to the dot-depth hierarchy and the
Straubing-Th\'{e}rien hierarchy were considered in
\cite{bacostth92,st94,mpt00,s04,cps06}. For more background on
regular languages, starfree languages, concatenation hierarchies,
and their decidability questions we refer to the survey articles
\cite{bro76,pin95,pin96b,pin96a,yu96,pinweil02,wei04}.

\begin{table}[t]   \label{pap_tbl_results}
\begin{center}
    \renewcommand{\arraystretch}{1.1}
    \begin{tabular}{p{5cm}|p{3cm}|p{4.5cm}}  
        Boolean hierarchy classes       & decidability          & complexity \\ \hline\hline
        $\lsig{1}(n)$                               & \cite{scwa98a}        & $\cNL$-complete \cite{scwa98a}\\ \hline
        ${\mathcal C}^1_k(n)$                   & \cite{gs01b,s01}          & $\cNL$-complete \cite{sch01}\\ \hline
        $\bsig{1}(n)$                               & \cite{gs01b}          & $\cNL$-complete [this paper]\\ \hline
        ${\mathcal C}^d_k(n)$               & \cite{s04}           & $\cNL$-complete [this paper] \\ \hline
        $\sigd_1(n)$                                &  \cite{s04}                                & $\cNL$-complete [this paper]\\ \hline
        $\sigt_1(n)$                                &  [this paper]                                & no efficient bound known\newline (see Remark~\ref{pap_rem_larged}) \\ \hline
        $\lsig{2}(n)$ for $|A|=2$       &  [this paper]         & $\cNL$-complete [this paper]
    \end{tabular}
    \vspace*{-3mm}
\end{center}
\caption{Overview of decidability and complexity results.}
\end{table}

{\bf Paper Outline. }
After the preliminaries,
we explain the general idea of an efficient membership algorithm
for the classes ${\mathcal C}^d_k(n)$ (section~\ref{pap_sec_intro_c}).
This easy example shows how a suitable characterization of a Boolean hierarchy
can be turned into an efficient membership test.
The algorithms for the other Boolean hierarchies are similar, but more complicated.
Section~\ref{pap_sec_new_char} provides new alternating-chain characterizations
for the Boolean hierarchies over $\bsig{1}$, $\sigd_1$, and $\lsig{2}$ for the alphabet $A=\{a,b\}$.
In section~\ref{pap_sec_decidability} we exploit these characterizations and
obtain efficient algorithms for testing the membership in these classes.
In particular, we obtain the decidability of the classes $\sigt_1(n)$ and $\lsig{2}(n)$ for the alphabet $A=\{a,b\}$.
Finally, section~\ref{pap_lower} provides lower bounds for the complexity of the
considered decidability problems.
As a consequence (with the exception of $\sigt_1(n)$) the membership problems
of all considered Boolean levels are logspace many-one complete for $\cNL$.
In contrast, the membership problems of the general classes
$\FO_\tau$ and $\FO_{\tau_d}$ are logspace many-one complete for $\cPSPACE$ and hence are strictly more complex.

Detailed proofs are available in the technical report \cite{gss07}.


\section{Preliminaries}

In this section we recall definitions and results that are needed later in the paper.
If not stated otherwise, $A$ denotes some finite alphabet with $|A|\geq 2$.
Let $A^\ast$ and $A^+$ be the sets of finite (resp., of finite non-empty)
words over $A$.
If not stated otherwise, variables range over the set of natural numbers.
We use $[m,n]$ as abbreviation for the interval $\{m, m+1, \ldots, n\}$.
For a deterministic finite automaton $M = (A, Z, \delta, s_0, F)$ (dfa for short),
the number of states is denoted by $|M|$ and the accepted language is denoted by $L(M)$.
Moreover, for words $x$ and $y$ we write $x \equiv_M y$ if and only if $\delta(s_0,x) = \delta(s_0,y)$.
For a class of languages ${\mathcal C}$, $\bc{{\mathcal C}}$ denotes the Boolean closure
of ${\mathcal C}$, i.e., the closure under union, intersection, and complementation.

All hardness and completeness results in this paper are with respect to logspace many-one reductions,
i.e., whenever we refer to $\cNL$-complete sets (resp., $\cPSPACE$-complete sets) then
we mean sets that are logspace many-one complete for $\cNL$ (resp., $\cPSPACE$).


\subsection{The Logical Approach to Regular Languages}\label{pap_prelogappr}


We relate to an arbitrary alphabet
$A=\{a,\ldots\}$ the signatures $\varrho=\{\leq, Q_a,\ldots\}$ and
$\sigma=\{\leq, Q_a, \ldots, \bot, \top, p, s\}$, where $\leq$ is a
binary relation symbol, $Q_a$ (for any $a\in A$) is a unary relation
symbol, $\bot$ and $\top$ are constant symbols, and $p,s$  are unary
function symbols. A word $u=u_0\ldots u_n\in A^+$ may be considered
as a structure ${\bf u}=(\{0,\ldots,n\};\leq,Q_a,\ldots)$ of signature
$\sigma$, where $\leq$ has its usual meaning, $Q_a(a\in A)$ are unary
predicates on $\{0,\ldots,n\}$ defined by $Q_a(i)\Leftrightarrow
u_i=a$, the symbols $\bot$ and $\top$ denote the least and the
greatest elements, while $p$ and $s$ are respectively the
predecessor and successor functions on $\{0,\ldots,n\}$ (with
$p(0)=0$ and $s(n)=n$). Similarly, a word $v=v_1\ldots v_n \in A^*$
may be considered as a structure ${\bf
v}=(\{1,\ldots,n\};\leq,Q_a,\ldots)$ of signature $\varrho$. For a
sentence $\phi$ of $\sigma$ (resp., $\varrho$), let $L_\phi=\{u\in
A^+|{\bf u}\models\phi\}$ (resp., $L_\phi=\{v\in A^*|{\bf
v}\models\phi\}$). Sentences $\phi,\psi$ are treated as equivalent
when $L_\phi=L_\psi$. A language is $\FO_\sigma$-definable (resp.,
$\FO_\varrho$-definable) if it is of the form $L_\phi$, where $\phi$
ranges over first-order sentences of $\sigma$ (resp., $\varrho$).
We denote by $\bsig{k}$ (resp., $\bpig{k}$) the class of languages that
can be defined by a sentence of $\sigma$
having at most $k-1$ quantifier alternations,
starting with an existential (resp., universal) quantifier.
$\lsig{k}$ and $\lpig{k}$ are defined analogously.
It is well-known that the class of $\FO_\sigma$-definable languages
(and $\FO_\varrho$-definable languages) coincides with the class
of {\em regular aperiodic languages} which are also known as the
{\em star-free languages}.
Moreover there is a levelwise correspondence to concatenation hierarchies:
The classes $\lsig{k}$, $\lpig{k}$, and
$\lbc{k}$ coincide with the classes of the Straubing-Th\'{e}rien
hierarchy \cite{pepi86}, while the classes $\bsig{k}$, $\bpig{k}$,
and $\bbc{k}$ coincide with the classes of the dot-depth hierarchy
\cite{tho82}.

We will consider also some enrichments of the signature $\sigma$.
Namely, for any positive integer $d$ let $\tau_d$ be the signature
$\sigma\cup\{P_d^0,\ldots,P_d^{d-1}\}$, where $P_d^r$ is the unary
predicate true on the positions of a word which are equivalent to
$r$ modulo $d$. By $\FO_{\tau_d}$-definable language we mean any
language of the form $L_\phi$, where  $\phi$ is a first-order
sentence of signature $\tau_d$. Note that signature $\tau_1$ is
essentially the same as $\sigma$ because $P_1^0$ is the valid
predicate.  In contrast, for $d>1$ the $\FO_{\tau_d}$-definable languages
need not to be aperiodic. E.g., the sentence $P_2^1(\top)$ defines the
language $L$ consisting of all words of even length which is known
to be non-aperiodic.  We are also interested in the signature $\tau=\bigcup_d\tau_d$.
Barrington et al.\ \cite{bacostth92,st94} defined quasi-aperiodic languages
and showed that this class coincides with the class of $\FO_{\tau}$-definable languages.
With the same proof we obtain the equality of the class of $d$-quasi-aperiodic languages
and the class of $\FO_{\tau_d}$-definable languages \cite{s04}.
It was observed in the same paper that $\sigt_n=\bigcup_d\sigd_n$ for
each $n>0$, where $\Sigma_n$ with an upper index denotes the class
of regular languages defined by $\Sigma_n$-sentences of the
corresponding signature in the upper index.

\begin{theorem} \label{pap_thm_decidibility_bc}
    For the following classes ${\mathcal D}$ it is decidable
    whether a given dfa $M$ accepts a language in ${\mathcal D}$:
    $\bc{\lsig{1}}$ \cite{simon75},
    $\bc{\lsig{2}}$ for $|A|=2$ \cite{str88},
    $\bc{\bsig{1}}$ \cite{kna83a},
    $\bc{\sigt_1}$ \cite{mpt00}.
\end{theorem}

We do not know the decidability of $\bc{\sigd_1}$.
However, it is likely to be a generalization of Knast's proof \cite{kna83a}.

\subsection{Preliminaries on the Classes $\boldsymbol{{\mathcal C}^d_k(n)}$}\label{prels}

We will also refer to `local' versions
of the BH's over $\bsig{1}$ and $\sigd_1$ \cite{stern85,gs01b,s01,s04}.
For any $k \ge 0$ the following partial order on $\sowp$ was studied in
\cite{stern85,gs01b,s01}: $u\leq_k v$, if $u=v\in A^{\leq k}$
or $u,v\in A^{>k}$, $p_k(u)=p_k(v)$,
$s_k(u)=s_k(v)$, and there is a $k$-embedding
$f:u\rightarrow v$. Here $p_k(u)$ (resp., $s_k(u)$) is the prefix (resp.,
suffix) of $u$ of length $k$, and the $k$-embedding $f$ is a
monotone injective function from $\{0.\ldots,|u|-1\}$ to
$\{0.\ldots,|v|-1\}$ such that $u(i)\cdots u(i+k)=v(f(i))\cdots
v(f(i)+k)$ for all $i<|u|-k$.
Note that $\leq_0$ is the subword relation.

\begin{definition}[\cite{s04}]\label{pap_kd}
    Let $k \ge 0$ and $d>0$.
    \begin{enumerate}
        \item We say that a $k$-embedding $f:u\rightarrow v$ is a
            $(k,d)$-embedding, if $P_d^r(i)$ implies $P_d^r(f(i))$ for all
            $i<|u|$ and $r<d$.
        \item  For all $u,v\in A^+$, let $u\leq^d_kv$ mean that $u=v\in
                A^{\leq k}$ or $u,v\in A^{>k}$, $p_k(u)=p_k(v)$, $s_k(u)=s_k(v)$,
                and there is a $(k,d)$-embedding $f:u\rightarrow v$.
        \item With ${\mathcal C}_k^d$ we denote  the class of all upper sets in $(A^+;\leq_k^d)$.
    \end{enumerate}
\end{definition}

Note that for $d=1$ the order $\leq^d_k$ coincides with $\leq_k$.
By an {\em alternating $\leq^d_k$-chain} of length $n$ for a set $L$ we mean a sequence $(x_0,\ldots,x_n)$
such that $x_0 \leq^d_k \cdots \leq^d_k x_n$ and $x_i\in L \Leftrightarrow x_{i+1}\not\in L$ for every $i<n$.
The chain is called {\em 1-alternating} if $x_0\in L$, otherwise it is called {\em 0-alternating}.

\begin{proposition}[\cite{gs01b,s01,s04}]\label{pap_lkd}
    For all $L\subseteq A^+$ and $n \ge 1$, $L\in{\mathcal C}_k^d(n)$
    if and only if $L$ has no 1-alternating chain of length $n$ in $(A^+;\leq_k^d)$.
\end{proposition}

Moreover, $(A^+;\leq_k^d)$ is a well partial order,
$\sigd_1=\bigcup_k{\mathcal C}_k^d$, and
$\sigt_1=\bigcup_{k,d}{\mathcal C}_k^d$ \cite{gs01b,s01,s04}.

\begin{theorem}[\cite{stern85}] \label{pap_thm_decidibility_bc_ck}
    It is decidable whether a given dfa  accepts a language in $\bc{{\mathcal C}^1_k}$.
\end{theorem}

For $d > 1$ it is not known whether $\bc{{\mathcal C}^d_k}$ is decidable.
However, we expect that this can be shown by generalizing the proof in \cite{stern85}.


\section{Efficient Algorithms for $\boldsymbol{{\mathcal C}^d_k(n)}$} \label{pap_sec_intro_c}

The main objective of this paper is the design of {\em efficient}
algorithms deciding membership for particular Boolean hierarchies.
For this, two things are needed: first, we need to prove suitable
characterizations for the single levels of these hierarchies. This
gives us certain criteria that can be used for testing membership.
Second, we need to construct algorithms that efficiently apply these
criteria. If both steps are successful, then we obtain an efficient
membership test.

Based on known ideas for membership tests for ${\mathcal C}^1_0(n)$
\cite{scwa98a}\footnote{For all $n$, the classes ${\mathcal C}^1_0(n)$ and $\lsig{1}(n)$
coincide up to the empty word, i.e., ${\mathcal C}^1_0(n) = \{L \cap A^+ \cond L \in \lsig{1}(n)\}$.}
and ${\mathcal C}^1_k(n)$ \cite{sch01}, in this section we explain the
construction of a nondeterministic, logarithmic-space membership algorithm
for the classes ${\mathcal C}^d_k(n)$.
This is the first efficient membership test for this general case.
Our explanation has an exemplary character,
since it shows how a suitable characterization of a Boolean hierarchy
can be turned into an efficient membership test.
Our results in later sections use similar, but more complicated constructions.

We start with the easiest case $k=0$ and $d=1$, i.e., with the classes ${\mathcal C}^1_0(n)$.
By Proposition~\ref{pap_lkd},
\begin{equation} \label{pap_eqn_01}
    \mbox{$L \notin {\mathcal C}^1_0(n)$ $\quad \gdw \quad$ $L$ has a 1-alternating $\le_0$-chain of length $n$.}
\end{equation}
We argue that for a given $L$, represented by a finite automaton $M$,
the condition on the right-hand side can be verified in nondeterministic logarithmic space.
So we have to test whether there exists a chain $w_0 \le_0 \cdots \le_0 w_n$ such that
$w_i \in L$ if and only if $i$ is even.
This is done by the following algorithm.

{\tt\small
    \begin{tabbing}
    \hspace*{10mm}\=
    \hspace*{7mm}\=\hspace*{6mm}\=
    \hspace*{6mm}\=\hspace*{6mm}\=
    \hspace*{6mm}\=\hspace*{6mm}\=
    \hspace*{6mm}\=\hspace*{6mm}\=
    \kill
    \>  0 \> // \parbox[t]{135mm}{
                    On input of a deterministic, finite automaton $\mathtt{M = (A, Z, \delta, z_0, F)}$\\
                    the algorithm tests whether $\mathtt{L(M)} \in {\mathcal C}^1_0(n)$.
                } \\
    \>  1 \> let $\mathtt{s_0 = \cdots = s_n = z_0}$ \\
    \>  2 \> do \\
    \>  3 \> \> nondeterministically choose $\mathtt{a \in A}$ and $\mathtt{j \in [0,n]}$ \\
    \>  4 \> \> for $\mathtt{i=j}$ to $\mathtt{n}$ \\
    \>  5 \> \> \> $\mathtt{s_i = \delta(s_i,a)}$ \hspace*{10mm} // stands for the imaginary command $\mathtt{w_i := w_i a}$\\
    \>  6 \> \> next $\mathtt{i}$ \\
    \>  7 \> until $\mathtt{\forall i, [s_i \in F \;\gdw\; \mbox{\tt $\mathtt{i}$ is even}]}$ \\
    \>  8 \> accept
    \end{tabbing}
}

The algorithm guesses the words $w_0, \ldots, w_n$ in parallel.
However, instead of constructing these words in the memory, it
guesses the words letter by letter and stores only the states $s_i =
\delta(z_0,w_i)$. More precisely, in each pass of the loop we choose
a letter $a$ and a number $j$, and we interpret this choice as
appending $a$ to the words $w_j, \ldots, w_n$. Simultaneously, we
update the states $s_j, \ldots, s_n$ appropriately. By doing so, we
guess all possible chains $w_0 \le_0 \cdots \le_0 w_n$ in such a way
that we know the states $s_i = \delta(z_0,w_i)$. This allows us to
easily verify the right-hand side of (\ref{pap_eqn_01}) in line~7.
Hence, testing non-membership in ${\mathcal C}^1_0(n)$ is in $\cNL$.
By $\cNL = \ccoNL$ \cite{imm88b,sze87}, the membership test also
belongs to $\cNL$.

The algorithm can be modified such that it works for ${\mathcal C}^d_0(n)$ where $d$ is arbitrary:
For this we have to make sure that the guessed $\le_0$-chain is even a $\le_0^d$-chain, i.e.,
the word extensions must be such that the lengths of single insertions are divisible by $d$.
This is done by (i) introducing new variables $l_i$ that count the current length of $w_i$ modulo $d$
and (ii) by making sure that $l_i = l_{i+1}$ whenever $j \le i < n$
(i.e., letters that appear in both words, $w_i$ and $w_{i+1}$,
must appear at equivalent positions modulo $d$).
So also the membership test for ${\mathcal C}^d_0(n)$ belongs to $\cNL$.

Finally, we adapt the algorithm to make it work for ${\mathcal C}^d_k(n)$ where $d$ and $k$ are arbitrary.
So we have to make sure that the guessed $\le_0^d$-chain is even a $\le_k^d$-chain.
For this, let us consider an extension $u \le_k^d w$ where $u,w \in A^{>k}$.
The $(k,d)$-embedding $f$ that is used in the definition of $u \le_k^d w$ ensures that
for all $i$ it holds that in $u$ at position $i$ there are the same $k+1$ letters as in $w$ at position $f(i)$.
Therefore, a word extension $u \le_k^d w$ can be split into a series
of elementary extensions of the form $u_1 u_2 \le_0^d u_1 v u_2$ such that
the length $k$ prefixes of $u_2$ and $v u_2$ are equal.
The latter is called the {\em prefix condition}.
Moreover, we can always make sure that the positions in $u$ at which the elementary extensions occur
form a strictly increasing sequence. This allows us to guess the words in the $\le_k^d$-chain letter by letter.
Now the algorithm can test the prefix condition by introducing new variables $v_i$
that contain a guessed preview of the next $k$ letters in $w_i$.
Each time a letter is appended to $w_i$,
(i) we verify that this letter is consistent with the preview $v_i$ and
(ii) we update $v_i$ by removing the first letter and by appending a new guessed letter.
In this way the modified algorithm carries the length $k$ previews of the $w_i$ with it
and it makes sure that guessed letters are consistent with these previews.
Moreover, we modify the algorithm such that whenever $j \le i < n$,
then the condition $v_i = v_{i+1}$ is tested.
The latter makes sure that elementary extensions $u_1 u_2 \le_0^d u_1 v u_2$
satisfy the prefix condition and hence the involved words are even in $\le_k^d$ relation.
This modified algorithm shows the following.

\begin{theorem} \label{pap_thm_ckd_in_NL}
    $\{M \cond M \mbox{ is a det.\ finite automaton and } L(M) \in {\mathcal C}^d_k(n) \} \in \cNL$ for
    $k \ge 0$, $d \ge 1$.
\end{theorem}

We now explain why the above idea does not immediately lead to a
nondeterministic, logarithmic-space membership algorithm for the
classes $\bsig{1}(n)$, although an alternating chain
characterization for $\bsig{1}(n)$ is known from \cite{gs01b}.
Note that the described
algorithm for ${\mathcal C}^d_k(n)$ stores the following types of
variables in logarithmic space.

\begin{enumerate}
    \item variables $s_i$ that contain states of $M$ 
    \item variables $l_i$ that contain numbers from $[0,d-1]$ 
    \item variables $v_i$ that contain words of length $k$
\end{enumerate}

However, the characterization of the classes $\bsig{1}(n)$ \cite{gs01b}
is unsuitable for our algorithm:
In order to verify the forbidden-chain condition,
we have to guess a chain of so-called structured words
and have to make sure that certain parts $u$ in these words are $M$-idempotent
(i.e., $\delta(s,u) = \delta(s,uu)$ for all states $s$).
Again we would try to guess the words letter by letter,
but now we have to make sure that (larger) parts $u$ of these words are $M$-idempotent.
We do not know how to verify the latter condition in logarithmic space.

In a similar way one observes that the known characterization of the classes
$\sigd_1(n)$ \cite{s04} cannot be used for
the construction of an efficient membership test.
So new characterizations of $\bsig{1}(n)$ and $\sigd_1(n)$ are needed
in order to obtain efficient membership algorithms.


\vskip-0.3cm
\section{New Characterizations of Boolean-Hierarchy Classes} \label{pap_sec_new_char}

In this section we develop new alternating-chain characterizations that allow the construction
of efficient algorithms deciding membership for the Boolean hierarchies over $\bsig{1}$, $\sigd_1$,
and $\lsig{2}$ for $|A|=2$.
We begin with the introduction of {\em marked words} and related partial orders
which turn out to be crucial for the design of efficient algorithms.


\subsection{Marked Words} \label{pap_subsec_markedwords}

For a fixed finite alphabet $A$, let $\mf{A}\eqdef \set{[a,u]}{a\in
A,u\in A^*}$ be the corresponding marked alphabet. Words over
$\mf{A}$ are called marked words. For $w\in\mf{A}^*$ with
$w=[a_1,u_1]\cdots[a_m,u_m]$ let $\unmark{{w}}\eqdef a_1\cdots
a_m\in A^*$ be the corresponding unmarked word.
Sometimes we use the functional notation $f_i(w)=a_1u^i_1\cdots a_mu^i_m$,
i.e., $f_0(w)=\unmark{{w}}$.
Clearly,
$f_0:\mf{A}^*\rightarrow A^\ast$ is a surjection.
For $x=x_1\cdots x_m\in A^+$ and $u\in A^*$ we define
$[x,u]\eqdef[x_1,\eps]\cdots[x_{m-1},\eps][x_m,u]$.

Next we define a relation on marked words.
For $w,w'\in\mf{A}^*$  we write $w \sw w'$ if and only if there exist
$m\geq 0$, marked words $x_i,z_i\in\mf{A}^*$, and marked letters
$b_i=[a_i,u_i]\in\mf{A}$ where $u_i\in A^+$ s.t.
%
\begin{eqnarray*}
    w &=& x_0 b_1 \;\; \phantom{z_1 b_1} \;\; x_1 b_2 \;\; \phantom{z_2 b_2} \;\; x_2 \;\;\;\; \cdots \;\;\;\; b_m \;\; \phantom{z_m b_m} \;\; x_m, \mbox{ and} \\
    w' &=& x_0 b_1 \;\; z_1 b_1 \;\; x_1 b_2 \;\; z_2 b_2 \;\; x_2 \;\;\;\; \cdots \;\;\;\; b_m \;\; z_m b_m \;\; x_m.
\end{eqnarray*}

We call $b_i$ the context letter of the insertion $z_i b_i$.
We write $w \sw^d w'$ if $w \sw w'$ and  $|f_0(z_i b_i)|\equiv 0$ (mod $d$) for all $i$.
Note that $\sw^1$ coincides with $\sw$ and
observe  that $\sw^d$ is a transitive relation.

For a dfa $M = (A, Z, \delta, s_0, F)$ and $s,t \in Z$
we write $s \con{w} t$, if $\delta(s,\overline{w})=t$ and for all $i$,
$\delta(s,a_1 \cdots a_i) = \delta(s,a_1 \cdots a_i u_i)$.
So $s \con{w} t$ means that the marked word $w$
leads from $s$ to $t$ in a way such that the labels of $w$
are consistent with loops in $M$.
We say that $w$ is $M$-consistent, if for some $t \in Z$, $s_0 \con{w} t$
and denote by ${\mathcal B}_M$ the set of marked words that are
$M$-consistent.
Every $M$-consistent word has the following nice property.

\begin{proposition} \label{pap_propo_consistent_equiv}
For $w=[c_1,u_1]\cdots[c_m,u_m]\in {\mathcal B}_M$ and all $j \ge 0$,
$f_0(w)\equiv_M   c_1u_1^j\cdots c_mu_m^j$.
\end{proposition}


\subsection{New Characterization of the Classes $\boldsymbol{\bsig{1}(n)}$ and $\boldsymbol{\sigd_1(n)}$} \label{pap_subsec_sigd1}

We extend the known characterization of the classes $\sigd_1(n)$ \cite{s04}
and add a characterization in terms of alternating chains on $M$-consistent
marked words. Because we can also restrict the length of the labels
$u_i$, we denote by  ${\mathcal B}_M^c$ for any $c>0$ the set of marked
words $[a_0,u_0]\cdots[a_n,u_n]$ that are $M$-consistent and satisfy
$|u_i|\leq c$ for all $i\leq n$.

\begin{theorem}\label{pap_sigdh1}
The following is equivalent for
$d,n\!\geq\! 1$, a dfa $M$, $c\!=\!|M|^{|M|}$, and $L\!=\!L(M)\!\subseteq\! A^+$.
\begin{enumerate}
    \item[(1)] $L\in\sigd_1(n)$
    \item[(2)] $f_0^{-1}(L)$ has no 1-alternating chain of length $n$ in
                    $({\mathcal B}_M;\sw^d)$
    \item[(3)]  $f_0^{-1}(L)$ has no 1-alternating chain of length $n$ in
        $({\mathcal B}^c_M;\sw^d)$
\end{enumerate}
\end{theorem}

%
%



The case $d=1$ is an alternative to
the known characterization of the classes $\bsig{1}(n)$ \cite{gs01b}.
\begin{theorem} \label{pap_b12_chain_christian}
    Let $M$ be a dfa, $L=L(M)\subseteq A^+$ and $n\geq 1$. Then
    $L\in\bsig{1}(n)$ if and only if $f^{-1}_0(L)$ has no 1-alternating chain of length $n$ in  $({\mathcal B}_M;\sw)$.
\end{theorem}



%


We can give an upper bound on $d$ for languages in $\sigt_1(n)$.
\begin{theorem}\label{pap_sigt}
For every dfa $M$, $c=|M|^{|M|}$, and $d=c!$,~~
$L(M)\in\sigt_1(n) \Rightarrow L(M)\in\sigd_1(n)$.
\end{theorem}


\subsection{Characterization of the Classes $\boldsymbol{\lsig{2}(n)}$ for $\boldsymbol{|A|=2}$} \label{pap_subsec_lsigtwo}

We obtain an alternating-chain characterization for the classes of the
Boolean hierarchy over $\lsig{2}$ for the case $|A|=2$.
This allows us to prove the first decidability result for this hierarchy.
Note that only in case $|A|=2$ decidability of $\bc{\lsig{2}}$ \cite{str88} and $\lsig{3}$ \cite{gs01} is known.


For $u\in A^*$ let $\alf{u}$ be the set of letters in $u$.
We say that a marked word
$w=[c_1,u_1]\cdots[c_m,u_m]$ satisfies the \textit{alphabet condition}
if for all $u_i\neq\epsilon$ it holds that $\alf{u_i}=A$.
\begin{theorem}\label{pap_b32_chain}
Let $A=\{a,b\}$, $n\geq 1$ and let $L(M)\subseteq A^*$ for some dfa $M$ such that $L=L(M)$ is a star-free language.
Then
 $L\in\lsig{2}(n)$ if and only if $f^{-1}_0(L)$ has no 1-alternating chain $(w_0,\ldots,w_n)$ in $({\mathcal B}_M;\sw)$
    such that all $w_i$ satisfy the alphabet condition.
\end{theorem}


\vskip-0.3cm
\section{Decidability and Complexity} \label{pap_sec_decidability}

The alternating-chain characterizations from the last sections can be used for the construction of efficient algorithms
for testing the membership in these classes.
As corollaries we obtain new decidability results:
the classes $\sigt_1(n)$ and $\lsig{2}(n)$ for $|A|=2$ are decidable.

The characterizations given in Theorems~\ref{pap_sigdh1} and \ref{pap_b12_chain_christian}
allow the construction of nondeterministic, logarithmic-space membership tests for $\bsig{1}(n)$ and $\sigd_1(n)$.

\begin{theorem} \label{pap_thm_efficient_membership_test}
    For all $n \ge 1$,
    $\{M \cond M \mbox{ is a det.\ finite automaton and }
    L(M) \in \bsig{1}(n) \} \!\in\! \cNL$.
\end{theorem}

\begin{theorem} \label{pap_thm_efficient_membership_test_sigd}
    For all $n \ge 1$,
    $\{M \cond M \mbox{ is a det.\ finite automaton and }
    L(M) \!\in\! \sigd_1(n) \} \!\in\! \cNL$.
\end{theorem}

\begin{remark} \label{pap_rem_larged}
\rm Unfortunately, we do not obtain $\cNL$-decidability for the
classes $\sigt_1(n)$. The reason is that the $d$ in
Theorem~\ref{pap_sigt} is extremely big, i.e., we only know the upper
bound $d \le (m^m)!$ where $m$ is the size of the automaton. We
leave the question for an improved bound open. Note that if $d$ can
be bounded polynomially in the size of the automaton, then
$\sigt_1(n)$ is decidable in $\cNL$. Although $d$ is very large, it
is still computable from the automaton $M$ which implies the
decidability of all levels $\sigt_1(n)$. This settles a question
left open in \cite{s04}.
\end{remark}

\begin{theorem} \label{pap_thm_membership_test_sigt}
    For all $n \ge 1$,  
   $\{M \cond M \mbox{ is a det.\ finite automaton and } L(M) \in \sigt_1(n) \}$ is decidable.
\end{theorem}

\begin{theorem} \label{pap_thm_membership_test_lsig2}
    For all $n \ge 1$,\\
    \centerline{$\{M \cond M \mbox{ is a det.\ finite automaton over the alphabet $\{a,b\}$ and } L(M) \in \lsig{2}(n) \} \in \cNL$.}
\end{theorem}


\vskip-0.3cm
\section{Exact Complexity Estimations} \label{pap_lower}

With the exception of $\sigt_1(n)$,
the membership problems of all classes of Boolean hierarchies considered in this paper
are $\cNL$-complete. In contrast, the membership problems of the general classes
$\FO_\tau$ and $\FO_{\tau_d}$ are $\cPSPACE$-complete and hence are strictly more complex.


\begin{proposition}\label{pap_low}
Let ${\mathcal C}$ be any class of regular quasi-aperiodic languages
over $A$ with $|A|\geq 2$ and $\emptyset\in{\mathcal C}$. Then it is $\cNL$-hard to
decide whether a given dfa $M$ accepts a language in ${\mathcal C}$.
\end{proposition}

Together with the upper bounds established in the previous sections
this immediately implies the following exact complexity estimations.

\begin{theorem} \label{pap_exact}
Let $k\ge 0$, $n \ge 1$,  $d \ge 1$ and ${\mathcal C}$ is one of the
classes ${\mathcal C}^d_k(n)$, $\bsig{1}(n)$, $\sigd_1(n)$, or
$\lsig{2}(n)$ for $|A|=2$.  Then $\{M \cond M \mbox{ is a
det.\ finite automaton and } L(M) \in {\mathcal C} \}$ is
$\cNL$-complete.
\end{theorem}

We conclude this section with a corollary of the
$\cPSPACE$-completeness of deciding $\FO_\sigma$ which was
established by Stern \cite{stern85b} and by Cho and Huynh
\cite{ch91}. It shows that the complexity of deciding the classes
$\FO_\tau$ and $\FO_{\tau_d}$ is strictly higher than the complexity
of deciding the classes mentioned in Theorem~\ref{pap_exact}. (Note that
$\cNL$ is closed under logspace many-one reductions,
$\cNL \subseteq \cDSPACE(\log^2 n)$ \cite{sav70a} and
$\cDSPACE(\log^2 n) \subsetneq \cPSPACE$ \cite{hs65}.
Hence the classes $\FO_\tau$ and $\FO_{\tau_d}$ can not be decided in $\cNL$.)

\begin{theorem}\label{pap_fo}
    The classes $\FO_\tau$ and $\FO_{\tau_d}$ are $\cPSPACE$-complete.
\end{theorem}


\vskip-0.3cm
\section{Conclusions}

The results of this paper (as well as several previous facts
that appeared in the literature) show that more and more
decidable levels of hierarchies turn out to be decidable in $\cNL$.
One is tempted to strengthen the well-known challenging conjecture of
decidability of the dot-depth hierarchy to the conjecture that all
levels of reasonable hierarchies of first-order definable regular
languages are decidable in $\cNL$. At least, it seems instructive to
ask this question about any level of such a hierarchy known to be
decidable.

In this paper we considered the complexity of classes of regular
languages only w.r.t.\ the representation of regular languages by
dfa's. Similar questions are probably open for other natural
representations of regular languages, like nondeterministic finite automata
and propositions of monadic second order, first order or temporal logics.


\vskip-0.3cm
\section*{Acknowledgements}
  We are grateful to Klaus W. Wagner for many helpful discussions.
  We would like to thank the anonymous referees for their valuable comments.


\newcommand{\etalchar}[1]{$^{#1}$}
\newcommand{\noopsort}[1]{}

\end{document}